\documentclass[slac_one]{revtex4}
\usepackage{graphicx}
\usepackage{fancyhdr}
\begin{document}
%
\title{HERA Combined Cross Sections and Parton Densities} 
\author{Burkard Reisert (on behalf of the H1 and ZEUS Collaborations)}
\affiliation{Max-Planck-Institut f\"ur Physik, F\"ohringer Ring 6, 80805 M\"unchen, Germany}
\begin{abstract}
~\\[-8mm]
 Deep inelastic scattering cross section measurements previously
published by the H1 and ZEUS collaborations are combined.
The procedure takes into account the systematic error correlations
in a coherent way, leading to a significantly reduced overall
cross section uncertainty.
The combined H1 and ZEUS measurements of the inclusive
neutral and charged current cross sections 
are used to perform a common NLO QCD fit.
The consistent treatment of systematic uncertainties
in the joint data set ensures that the resulting set of parton density
functions (PDFs) have a much reduced experimental uncertainty
compared to the previous PDF extractions performed separately
by the H1 and ZEUS collaborations. 
~\\[-6mm]
\end{abstract}
\maketitle
\thispagestyle{fancy}
\section{Introduction}
~\\[-9mm]
Deep inelastic scattering (DIS) of leptons off nucleons has been essential for our understanding 
of the structure of the nucleon. Colliding electrons and positrons with protons, HERA at DESY 
has played a unique role in revealing details of the proton substructure of quarks and gluons. 
In June 2007 the operation of HERA was terminated after a period of 15 years of data taking.
During this time the H1 and ZEUS collaborations at HERA successfully operated their general purpose 
detectors which were adapted in 2001 to the luminosity upgrade of the collider interaction 
regions. Both collaborations 
have published accurate measurements of $e^\pm p$ DIS cross sections
based on about 120~$pb^{-1}$ of $ep$ data collected until 2000~\cite{H1:data:1,H1:data:2,H1:data:3,
H1:data:4,ZEUS:data:1,ZEUS:data:2,ZEUS:data:3,ZEUS:data:4,ZEUS:data:5,ZEUS:data:6}. These measurements
have been crucial in substantially constraining the parton density functions (PDFs) of the proton 
in global next-to-leading-order (NLO) QCD fits~\cite{MRST,CTEQ,ZEUS:global}, which use fixed target 
data as well as HERA data. The H1 and ZEUS collaborations have both demonstrated their capability
to extract complete sets of PDFs from their data alone~\cite{H1:data:4,ZEUS:alone} achieving a precision 
competitive to that obtained in the global QCD fits. 
Aiming for the most accurate determination of PDFs, H1 and ZEUS are now working towards combining 
their datasets and joining their QCD analyses.   
~\\[-10mm]
\section{Cross Sections, Structure Functions and Parton Distributions}
~\\[-9mm]
The kinematics of lepton hadron scattering is described in terms of the 
variables $Q^2$, the four-momentum transfer squared of the exchanged 
vector boson, Bj{\o}rken $x$, the fraction of the momentum of the incoming nucleon
carried by the struck quark (in the quark-parton-model, QPM), and inelasticity $y$, which is a measure of the energy transfered between the lepton and the nucleon.  
The differential cross section for the neutral current (NC) process, 
$e^\pm p\rightarrow e^\pm X$, is 
given in terms of the structure functions by 
\begin{equation}\label{eq:xsec:nc}
\frac{d^2\sigma^\pm_{NC}}{dxd Q^2} = 
\frac{2\pi\alpha^2}{xQ^4}
(Y_+\tilde{F}_2\mp Y_-x\tilde{F}_3 - y^2\tilde{F}_L),
\end{equation}
where $\alpha$ is the fine structure constant, $Y_\pm = 1\pm (1-y)^2$, and 
 $\tilde{F}_2$, $x\tilde{F}_3$ 
and $\tilde{F}_L$ denote the generalized NC proton structure functions 
which include $Z$ boson exchange contributions.
%
The structure functions $\tilde{F}_2$, $x\tilde{F}_3$  are directly related 
to quark distributions, and their $Q^2$ dependence is predicted by perturbative
QCD.  The dominant contribution to the cross section is $\tilde{F}_2$, which in turn is 
dominated by the electromagnetic structure function $F_2$.
In the QPM, $F_2$ -- below the bottom mass threshold -- can be written as 
\begin{equation}\label{eq:sfun:nc}
F_2 = \frac{4}{9}\left[x (u+c+\overline{u}+\overline{c})\right] + \frac{1}{9}\left[x(d + s +\overline{d}+\overline{s})\right].
\end{equation}
For low $x$, $x\leq 10^{-2}$, $F_2$ is dominated by the sea quarks, 
while its $Q^2$ evolution is controlled by the gluon contribution.
Thus HERA data provide crucial information on the low-$x$ sea-quark 
and gluon distributions. At high $Q^2$,
the structure function $x\tilde{F}_3$ becomes increasingly important, 
and provides information on the valence quark distributions, $u_v = u-\overline{u}$ and 
$d_v = d-\overline{d}$.
 
The charged current (CC) interactions,  
$e^+(e^-) p\rightarrow \overline{\nu}(\nu)X$,
provide additional flavor-type specific information to separate 
quark and anti-quark distributions at high-$x$, since their (leading-order) cross sections are given by
\begin{eqnarray}\label{eq:xsec:cc}
\frac{d^2\sigma^+_{CC}}{dxdQ^2} & = & 
\frac{G^2_F}{2\pi x}\left[\frac{M_W^2}{Q^2+M_W^2}\right]^2
x\left[(\overline{u}+\overline{c})+(1-y)^2(d+s)\right],\\
\frac{d^2\sigma^-_{CC}}{dxdQ^2} & = &
\frac{G^2_F}{2\pi x}\left[\frac{M_W^2}{Q^2+M_W^2}\right]^2
x\left[(u+c)+(1-y)^2(\overline{d}+\overline{s})\right],
\end{eqnarray}
with $G_F$ being the Fermi coupling constant and $M_W$ the mass of the 
exchanged W boson. 

The $e^\pm p$ NC and CC cross sections can thus be written completely 
in terms of up-type, $xU=x(u+c)$, down-type, $xD=x(d+s)$, and of  
anti-quark type, $x\overline{U}=x(\overline{u}+\overline{c})$ and $x\overline{D}=x(\overline{d}+\overline{s})$),
 distributions. These (anti-)quark type distributions
can be determined from HERA data alone. This has the advantage that there is 
no need for heavy target corrections, which must be applied to the 
$\nu {\rm Fe}$ and $\mu D$ fixed target data used in the global fits. Furthermore,
there is no need to assume isospin symmetry, i.e. that $d$ in the proton is the same 
as $u$ in the neutron, since information on the $d$ distribution can be obtained directly from CC $e^+p$ data.
~\\[-8mm]
\section{Combination of H1 and ZEUS Cross Sections}\label{sec:xsec:combine}
\begin{figure}
\includegraphics[width=0.45\textwidth, height=0.44\textwidth]{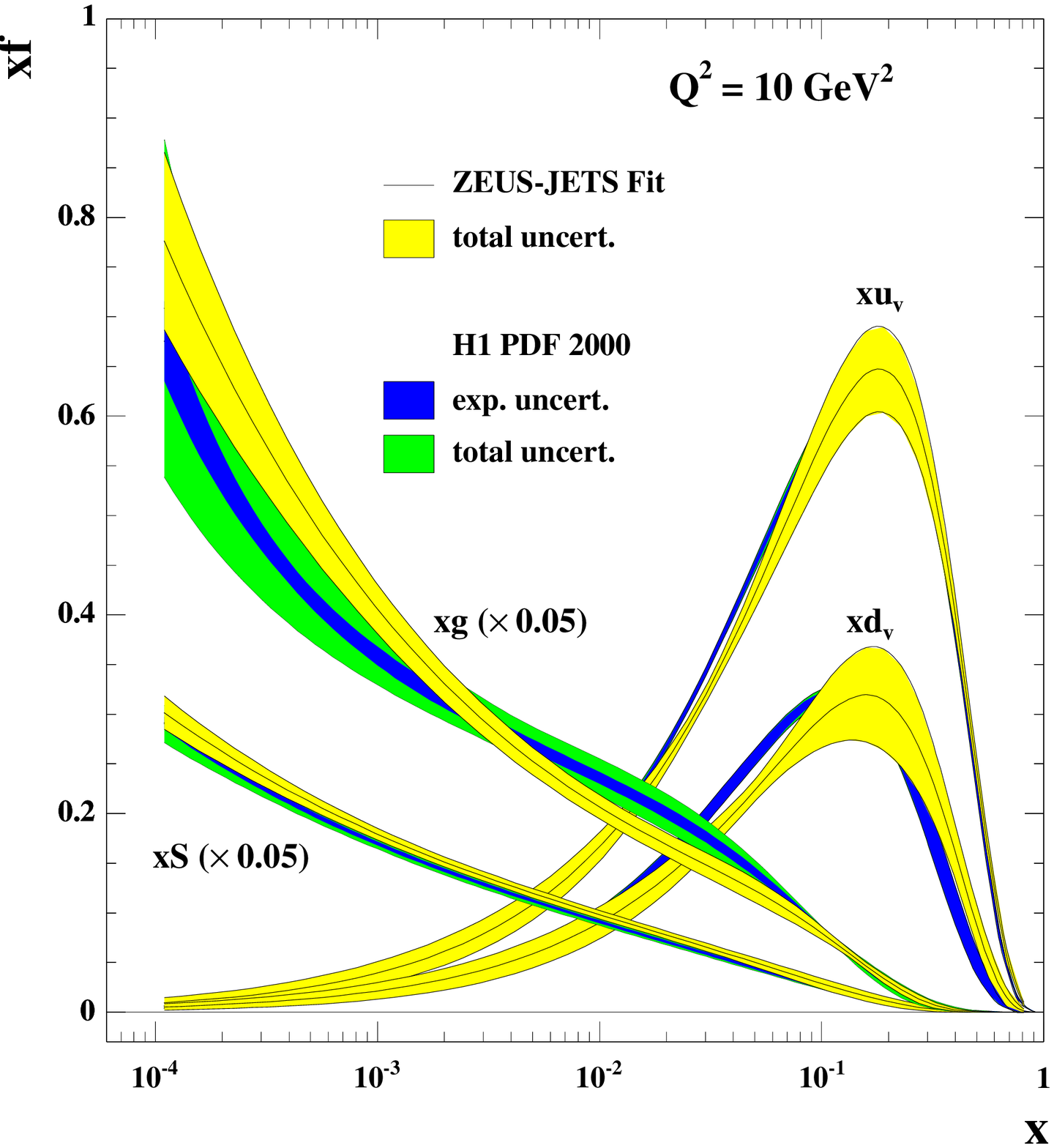}
\put(-190,200){(a)}%
\includegraphics[width=0.5\textwidth, height=0.49\textwidth]{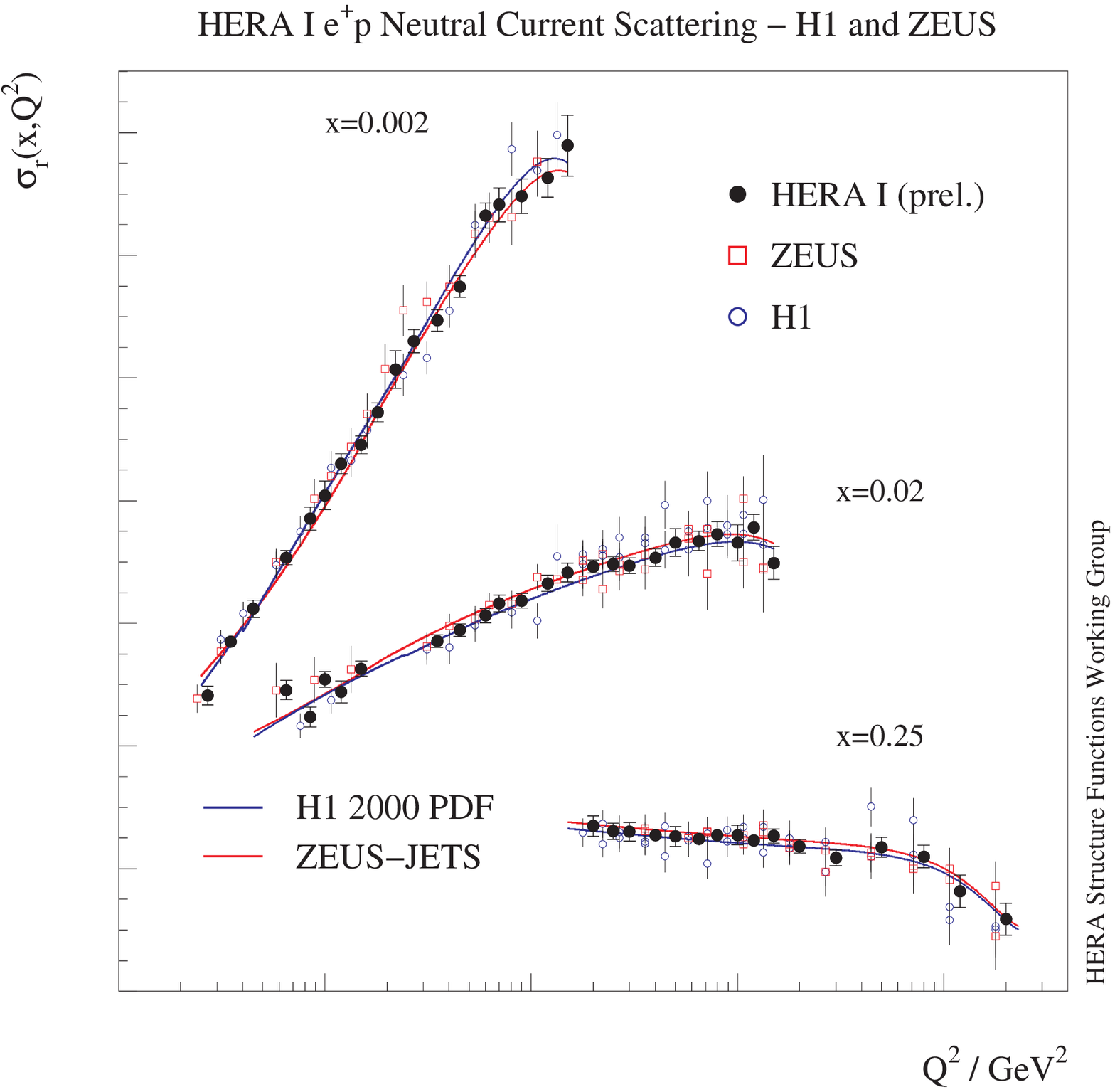}
\put(-215,200){(b)}%
\caption{Comparison of the PDFs and their uncertainties from ZEUS and H1 individual QCD analyses 
at a fixed value of $Q^2=10$~GeV$^2$ (a) and 
(b) neutral current $e^+p$ cross sections measurements for three selected $x$ bins 
as function of $Q^2$. The H1 (open points) and ZEUS data (open squares) are compared to 
the H1 and ZEUS combined data (full points). Measurements from the individual experiments 
have been shifted for clarity. The error bars show the total uncertainty. The curves are NLO QCD fits 
as performed by H1 and ZEUS to their own data.\label{fig:xsec}}
\end{figure}

~\\[-8mm]
The H1 and ZEUS collaborations have both used their data to perform PDF 
fits~\cite{H1:data:4,ZEUS:alone}. Both these data sets have small statistical errors, 
so that the contributions of systematic uncertainties become increasingly important 
and their proper treatment is essential. Basically there are two approaches, the 
Offset method used in the ZEUS analysis and the Hessian method applied by the H1 analysis. 
The resulting ZEUS and H1 PDFs are compatible, although the gluon PDFs do have somewhat 
different shapes, as is shown in Fig.~\ref{fig:xsec}a.

It is possible to improve on this situation by averaging the H1 and ZEUS 
datasets in a model-independent way prior to performing a QCD analysis 
on them. Since both experiments measure the same cross 
section at a given kinematic point one can minimize the following $\chi^2$ and
obtain a 
common true value at any measured kinematic point together with the correlated systematic shifts 
\begin{equation}\label{eq:chi2:combine}
\chi^2_{exp}(M^{i,true},\Delta\alpha_j) = 
\sum_i \frac{\left[M^{i,true} - \left(M^i + \sum_j\frac{\partial M^i}{\partial\alpha_j}\Delta\alpha_j\right)\right]^2}{\delta^2_i} + \sum_j\frac{\Delta\alpha^2_j}{\delta^2_{\alpha_j}}.  
\end{equation}
Here $M^i$ is the measured central value, and $\delta_i$ the statistical and 
uncorrelated systematic uncertainty of the quantity $M$. 
The $M^{i,true}$ are the values following from the minimization; 
$\Delta\alpha_j$ are parameters for the $j^{th}$ source of correlated systematic uncertainty and $\partial M^i/\partial\alpha_j$ denotes the sensitivity of point $i$ to source $j$. For the cross section measurements the index $i$  labels a particular measurement at a given $(x, Q^2)$. Equation~\ref{eq:chi2:combine} represents the correlated probability distribution function for the quantity $M^{i,true}$ and for the systematic uncertainty $\Delta\alpha_j$. 

The $\chi^2$ defined in  Eq.~\ref{eq:chi2:combine} is suitable for 
measurements in which the uncertainties are absolute or {\it additive}, 
i.e. do not depend on the central value of the measurement. 
For the cross section measurement, however, the correlated and uncorrelated 
systematic errors are proportional to the central values. This proportionality 
can be approximated by a linear dependence. In this case the combination 
of the data sets using Eq.~\ref{eq:chi2:combine}
has a bias towards lower cross section values since the measurements with 
small central values have smaller absolute uncertainties. An improved 
$\chi^2$ can be defined by replacing $\delta_i \rightarrow \frac{M^{i,true}}{M^i}\delta_i$ and$\frac{\partial M^i}{\partial\alpha_j}\Delta\alpha\rightarrow \frac{\partial M^i}{\partial\alpha_j}\frac{M^{i,true}}{M^i}\Delta\alpha$ which 
translates the relative or {\it multiplicative} uncertainties for each measurement to the absolute uncertainty. 

The correlated systematic uncertainties  are floated coherently such that 
each experiment calibrates the other one. This allows a significant reduction 
of the correlated systematic uncertainty for much of the kinematic plane, 
as is shown in Fig.~\ref{fig:xsec}b for three representative $x$-bins.

A study of the global $\chi^2/NDF$ of the average and of the pull 
distribution provides a model-independent consistency check between the 
experiments. The H1 and ZEUS data are found to be consistent, thus allowing 
to calculate the experimental uncertainties of the PDFs using 
the $\chi^2$ tolerance, $\Delta\chi^2 = 1$. This represents a further 
advantage compared to those global fits where increased tolerances 
of $\Delta \chi^2 \gg 1$ had to be introduced to ensure that all input 
data sets are consistent with the result of the global fit 
at a 90\% confidence level.
\section{QCD Analysis}\label{sec:qcd:ana}
~\\[-10mm]
QCD predictions for the structure functions are obtained by solving the DGLAP evolution 
equations at  NLO in the ${\rm\overline{MS}}$ scheme with the renormalization and factorization
scales chosen to be $Q^2$. The DGLAP equations yield the PDFs at all values of $Q^2$ provided 
they are input as functions of $x$ at some input scale $Q^2_0$. 
For our central fit we parameterize the gluon, $xg$, the valence quarks, $xu_v$ and $xd_v$,
as well as the sea anti-quark type $x\overline{U}$ and $x\overline{D}$ using the
generic form  
\begin{equation}
xf_i(x) = A_ix^{B_i}(1-x)^{C_i}\left(1+D_ix+E_ix^2+F_ix^3\right) \quad {\rm~with~} i=g, u_v, d_v, 
\overline{U}, \overline{D} 
\end{equation}
and the number of parameters is chosen by saturation of the $\chi^2$, 
such that parameters $D_i$, $E_i$, $F_i$ are  varied, i.e. they are allowed to take values 
different from 0 during the fit, but are incorporated only 
if there is a significant improvement of the $\chi^2$.
This leads us to set the parameters $D_i$, $E_i$, $F_i = 0$, for all partons except $xu_v$ for
which only $F_{u_v}=0$. The normalization parameters $A_{u_v}$ and $A_{d_v}$ are constrained 
to impose the valence quark number sum-rules and  $A_g$ is constrained to impose the 
momentum sum-rule. The $B_i$ parameters which determine the low-$x$ behavior of the PDFs are 
constrained such that there is a single $B$ parameter for the valance quark distributions $B_{u_v} = B_{d_v} \equiv B_v$ 
and another $B$ parameter for the sea anti-quark type distributions, 
$B_{\overline{U}} = B_{\overline{D}} \equiv B_{\overline{Q}}$. A flavor decomposition of the 
(anti-) quark type into individual quark flavors can be achieved by assuming that the strange and charm
quark at the input scale can be expressed as $x$ independent 
fractions, $f_s$ and $f_c$, of $\overline{D}$ and $\overline{U}$ respectively. 
Imposing that $\overline{d}-\overline{u}\rightarrow 0$ as 
$x\rightarrow 0$, a constraint implicit to global fits, further 
constrains $A_{\overline{U}}=A_{\overline{D}}\cdot(1-f_s)/(1-f_c)$.
The value of $f_s=0.33$, has been chosen to be consistent with determinations of the strange fraction
using neutrino-induced di-muon production. 
The charm fraction, $f_c=0.15$, has been set to be consistent with dynamic generation 
of the charm from the starting point of $Q^2=m_c^2$ in a zero-mass-variable-flavor-number scheme.
In total 11 parameters of the PDFs at an input scale $Q_0^2=4$~GeV$^2$ are obtained from the fit.
\section{Results}
\begin{figure}[t]
\includegraphics[width=0.45\textwidth, height=0.35\textwidth]{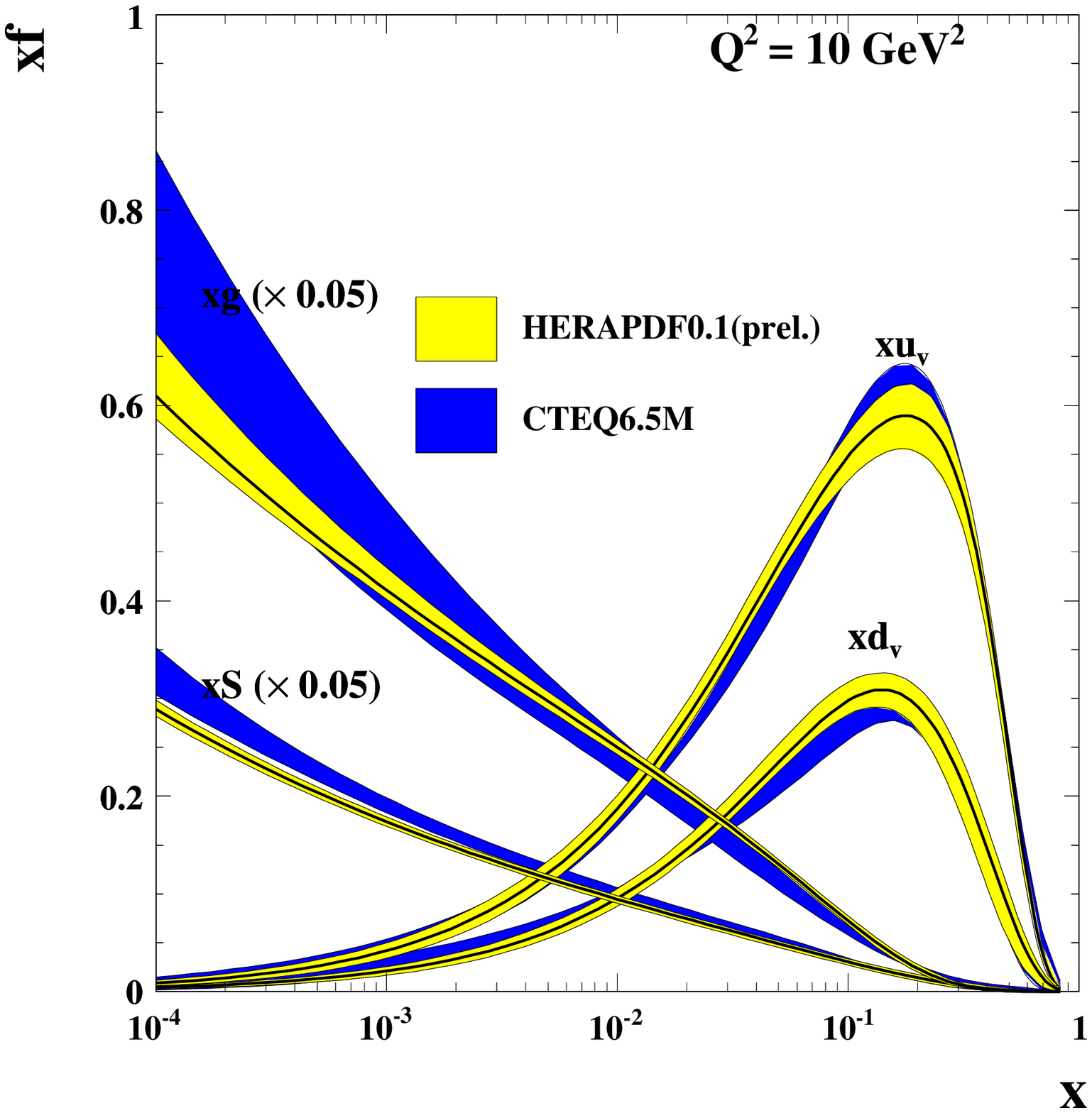}%
\put(-180,160){(a)}%
\includegraphics[width=0.45\textwidth, height=0.35\textwidth]{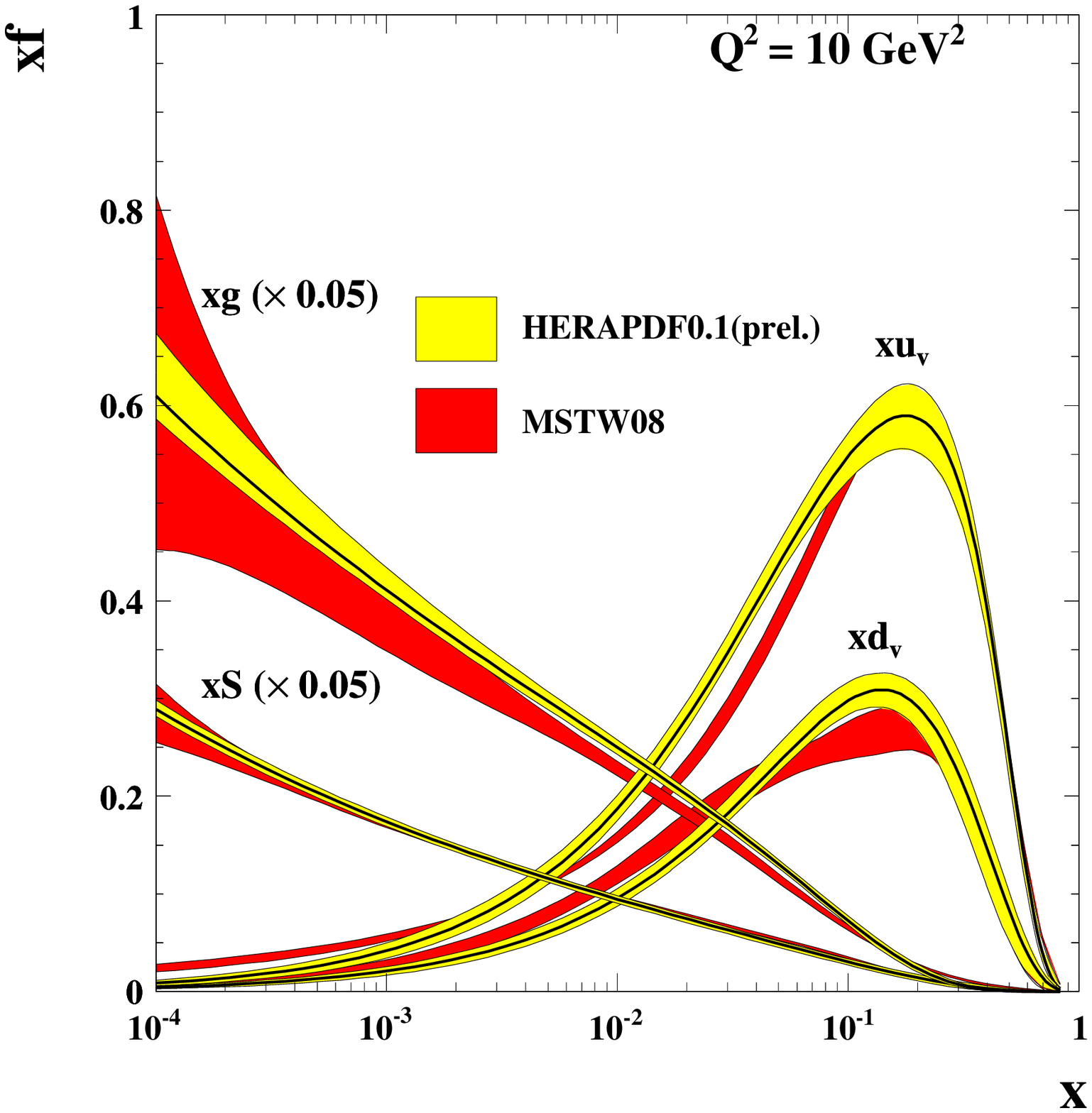}%
\put(-180,160){(b)}%
~\\[-5mm]
\caption{PDFs at $Q^2=10$~GeV$^2$ from the H1 and ZEUS combined data, 
HERAPDF0.1, compared to recent global PDF analyses, CTEQ6.5M~\cite{CTEQ} (a) and 
MSTW08~\cite{MSTW08} (b).\label{fig:pdfs}\\[-6mm]}
\end{figure}

~\\[-8mm]
The set of PDFs obtained from a NLO QCD analysis  performed on the combined data set 
of $e^\pm p$ NC and CC cross sections, discussed above, is shown in
 Fig.~\ref{fig:pdfs} (labeled HERAPDF0.1). The much reduced experimental uncertainties of the
combined cross sections propagate to the error bands of the PDFs, which also include six sources 
of model uncertainties due to variation of $m_c$, $m_b$, $f_s$, $f_c$, $Q_0^2$, and $Q_{min}^2$, 
the minimum $Q^2$ of the data included in the fit. The central fit, which in addition to all 
systematic sources of the 
combined data set, takes into account 4 sources of uncertainties from the combination procedure
achieves an excellent $\chi^2/NDF$ of 477/562. Figure~\ref{fig:pdfs} also 
compares the HERAPDF0.1 to recent global fits. 
Note that the HERAPDF0.1 results employs new, still preliminary HERA data of improved 
accuracy as discussed above. A further difference to the global fits is that these involve
data from a much larger variety of experiments and  physics processes  
and thus are forced to use an error definition 
unlike the~$\Delta \chi^2=1$~criterion applicable to the HERA data alone. 

The preliminary PDF set presented here is obtained from the combined,
almost complete set of cross section measurements 
of the pre-upgrade data taking period. 
The current PDF set is available at the LHAPDF library and will be published soon, 
when the HERA I neutral current cross section data are released. The data and QCD analyses
thus provide a valuable input for the physics at the Tevatron and the LHC. 
~\\[-10.5mm]

\end{document}